# Similarities Between the Inner Solar System and the Planetary System of PSR B1257+12

**Tsevi Mazeh** and **Itzhak Goldman**

School of Physics and Astronomy, Sackler Faculty of Exact Sciences
Tel Aviv University, Tel Aviv 69978, Israel

**Abstract** We call attention to the surprising similarity between the newly discovered planetary system around PSR B1257+12 and the inner solar system. The similarity is in the ratios of the orbital radii and the masses of the three planets.

The existence of planets orbiting the millisecond pulsar PSR B1257+12 (Wolszczan & Frail 1992) has been recently confirmed by Wolszczan (1994). This is the first convincing case for an extra-solar planetary system. The system was shown to include three planets with masses of $2.8 M_\oplus / \sin i_1$, $3.4 M_\oplus / \sin i_2$, and $0.015 M_\oplus / \sin i_3$, respectively. The angles denote the inclinations of the planets orbits with respect to our line of sight. Wolszczan (1994) has shown that the angles $i_1$ and $i_2$ are close to 90°, and coplanarity of the three planetary orbits would imply the same for $i_3$. The corresponding distances from the pulsar are 0.47, 0.36, and 0.19 AU, respectively (Wolszczan 1994).

We wish to draw attention to two striking similarities between the pulsar planetary system and the inner solar system *including the innermost three planets*. The first similarity involves the relative masses in the two systems. As known, the masses of the innermost three planets in the solar system are (Allen 1973) $1 M_\oplus$, $0.815 M_\oplus$ and $0.055 M_\oplus$ for Earth, Venus and Mercury, respectively. Thus, both systems include two outer planets with similar masses, with differences less than 20%, and an inner planet of a substantially smaller mass. Moreover, the masses of the two outer planets in both systems are of the same order of magnitude.

Even more impressing is the similarity between the distances of the planets to the central star, when normalized by the distance of the outermost third planet. In the pulsar case the normalized distances are 1, 0.77 and 0.40, while the solar values are (Allen 1973) 1, 0.72, and 0.39. The two sets of normalized distances are identical up to about 5%.

To illustrate the above similarities we show in Figure 1 the masses and the normalized distances from the central star, for the two planetary systems. Also indicated are the solar and the pulsar mass (taken to be 1.4 $M_\odot$), which on the scale of the figure occupy the same position. By construction, the outermost planet in each system is at a normalized distance of unity. To appreciate the similarity of the two systems, one should note that the masses shown in the figure span a range of more than seven orders of magnitude. The similarity in the distances ratios is reflected by the fact that the normalized distances of the two inner planets are almost equal for both systems.

The similarity between the two systems is outstanding, considering the differences in their evolutionary history. The solar planets are believed to have formed in the protoplanetary solar accretion disk (see, e.g., Cameron 1988), while the pulsar planets probably have evolved in a disk consisting of the debris of the binary companion of the millisecond pulsar (see, e.g., Podsiadlowski 1993 and Phinney & Hansen 1993). Understandably, one should be very careful in drawing conclusions based only on two systems. The assessment of the similarity significance must await the discovery of more planetary systems. Nevertheless, the similarity between the only two known systems, notably the almost identical distance ratios, may point to a possible global underlying mechanism of formation for both systems.

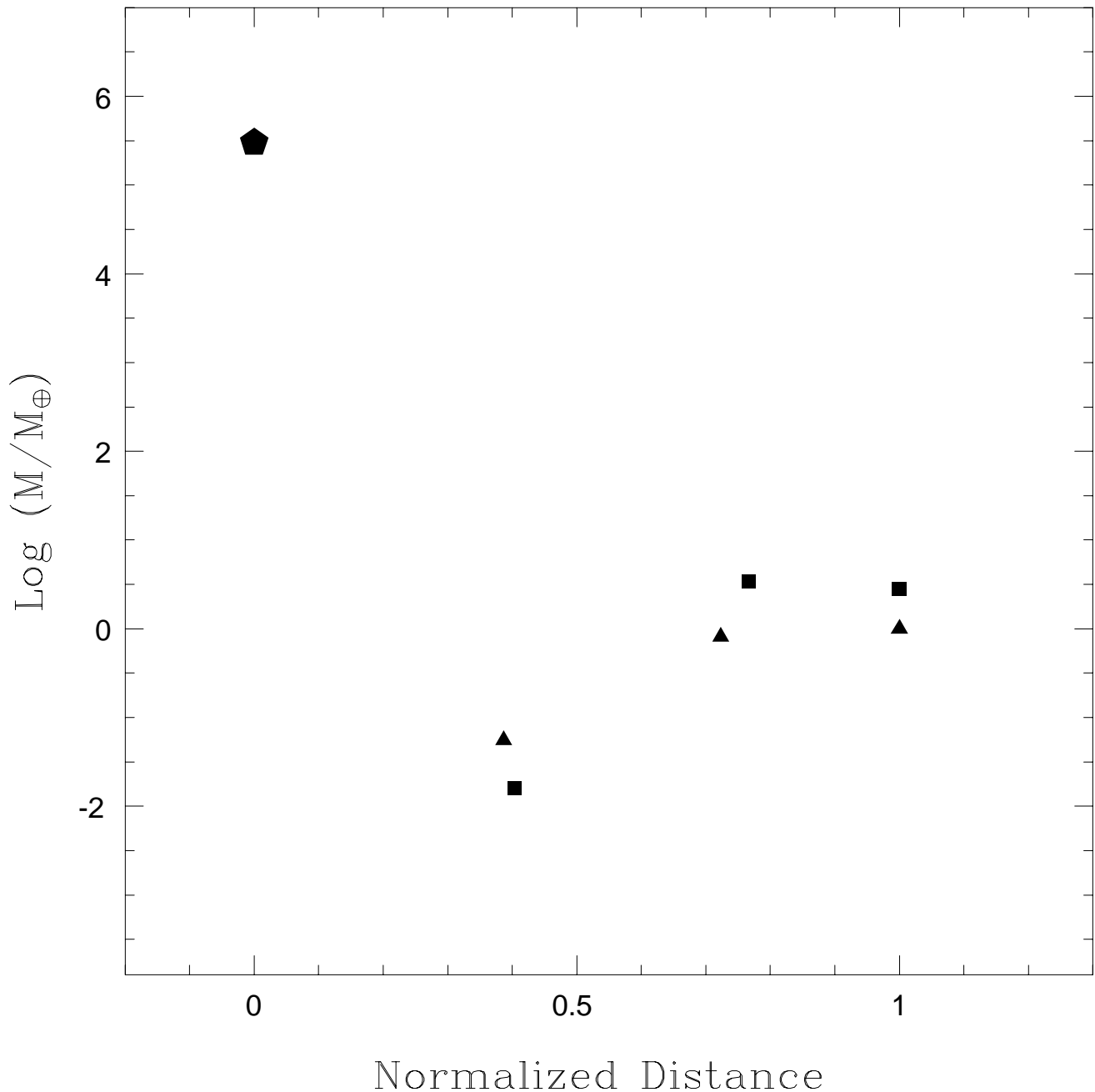

Figure 1: The masses and normalized distances of the three planets in the two systems. The unit mass is the Earth mass. The triangles represent the solar system planets and the squares the planets of the pulsar. The sun and the pulsar are represented by the large pentagon, at zero distance. On the scale of the figure, the position of the pulsar (mass of 1.4 $M_\odot$) coincides with that of the sun.

4